\documentclass[conference]{IEEEtran}
\IEEEoverridecommandlockouts
\usepackage[latin1]{inputenc} 
\usepackage{cite}
\usepackage{amsmath,amssymb,amsfonts}
\usepackage{graphicx}
\usepackage{textcomp}
\usepackage{algpseudocode}
\usepackage{algorithm} 
\usepackage{multirow}
\usepackage{tabularx}
\usepackage{dblfloatfix}
\usepackage{color}
\usepackage{xcolor}
\usepackage{subcaption}
\usepackage{algorithm}
\usepackage{algcompatible}
\usepackage{comment}
\usepackage{csquotes}
\usepackage{multicol}
\usepackage{booktabs}
\usepackage{gensymb}
\usepackage{forest}
\usepackage{tikz-qtree}
\usepackage{cleveref}
\usepackage{tikz}
\usetikzlibrary{shapes,arrows}
\usepackage{commath}
\usepackage{pbalance}
\usepackage[labelformat=simple]{caption}

\usepackage{algpseudocode}

\algnewcommand\algorithmicswitch{\textbf{switch}}
\algnewcommand\algorithmiccase{\textbf{case}}
\algnewcommand\algorithmicassert{\texttt{assert}}
\algnewcommand\Assert[1]{\State \algorithmicassert(#1)}

\def\BibTeX{{\rm B\kern-.05em{\sc i\kern-.025em b}\kern-.08em
    T\kern-.1667em\lower.7ex\hbox{E}\kern-.125emX}}

\title{

iThermTroj: Exploiting Intermittent Thermal Trojans in Multi-Processor System-on-Chips

\thanks{The work is partially supported by the National Science Foundation under grants number 2219679 and 2219680.
}
}

\begin{document}

 \author{\IEEEauthorblockN{Mehdi Elahi$^1$, Mohamed R. Elshamy$^2$, Abdel-Hameed Badawy$^2$, Ahmad Patooghy$^1$}
 $^1$ Department of Computer Systems Technology, North Carolina A\&T State University, NC, 27411\\
$^2$ New Mexico State University, United states\vspace{-4mm}
 }
 \maketitle

\begin{abstract} 
Thermal Trojan attacks present a pressing concern for the security and reliability of System-on-Chips (SoCs), especially in mobile applications. The situation becomes more complicated when such attacks are more evasive and operate sporadically to stay hidden from detection mechanisms. In this paper, we introduce Intermittent Thermal Trojans (iThermTroj) that exploit the chips' thermal information in a random time-triggered manner. According to our experiments, iThermTroj attack can easily bypass available threshold-based thermal Trojan detection solutions. We investigate SoC vulnerabilities to variations of iThermTroj through an in-depth analysis of Trojan activation and duration scenarios. We also propose a set of tiny Machine Learning classifiers for run-time anomaly detection to protect SoCs against such intermittent thermal Trojan attacks. Compared to existing methods, our approach improves the attack detection rate by 29.4\%, 17.2\%, and 14.3\% in scenarios where iThermTroj manipulates up to 80\%, 60\%, and 40\% of SoC's thermal data, respectively. Additionally, our method increases the full protection resolution to 0.8 degrees Celsius, meaning that any temperature manipulations exceeding $\pm 0.8$ degrees will be detected with 100\% accuracy.

\end{abstract}


\vspace{-1mm}
\section{Introduction}
\label{sec:Intro}

Demand for high-performance mobile systems equipped with multi-core processors is continually on the rise \cite{jeong2022band,10356076,10765816}. Mobile System-on-Chips (SoCs) have emerged as the backbone of these systems, catering to a myriad of tasks integral to the human's daily life. However, despite the ever-increasing performance advancements, mobile SoCs face a significant constraint in the form of a limited thermal-power budget \cite{wang2023efficient}. This limitation is primarily attributed to the challenges posed by maintaining compact form factors and limited cooling capabilities essential for handheld devices like smartphones \cite{8509120,info16030225}. Ensuring user comfort by mitigating skin temperature to prevent discomfort or burning sensations adds another layer of complexity to the thermal management puzzle \cite{gong2023performance}. 
Hence, Dynamic Thermal Management (DTM) techniques have been widely utilized among mobile systems to effectively regulate and minimize high operating temperatures \cite{4510751}.

The accuracy and reliability of thermal sensors (that DTM methods rely on) are crucial for effective thermal management strategies \cite{ding2020novel}. Thermal sensor malfunctions can occur due to a variety of factors, ranging from unintentional faults to deliberate tampering. Unintentional faults may include issues arising from fabrication defects, aging effects, susceptibility to noise, and process variations \cite{Oukaira2022FEMbasedTP,jlpea4040304,patooghy2019your}. These factors can lead to inaccuracies in thermal measurements, which can, in turn, impact the overall performance and reliability of the chip. 
Additionally, there are significant security risks associated with deliberate tampering, such as the insertion of Hardware Trojans (HT), which are malicious modifications trying to alter the behavior of thermal sensors, leading to falsified temperature values \cite{hasegawa2023node}. Erroneous temperature readings can trigger unnecessary frequency reduction or core throttling by DTM, which may reduce the performance of the chip. Furthermore, incorrect thermal data can accelerate the aging process of the chip, thereby reducing its lifespan. The insertion of thermal HTs thus poses a significant threat to the security and reliability of thermal management systems.

To address these challenges, several strategies have been proposed in the literature. Authors in \cite{abdelrehim2022bic} introduce the Blind Identification Countermeasure (BIC), a technique derived from the Blind Power Identification (BPI) algorithm \cite{said2018understanding}. BIC aims to detect, contain, and isolate malicious sensors, thereby offering accurate temperature estimations and safeguarding chip integrity. Through extensive testing across multiple processor architectures, the efficacy and accuracy of BIC are demonstrated, marking a significant advancement in enhancing the reliability and performance of mobile SoCs amidst evolving security threats and thermal management challenges. The contributions of this paper are as follows.
\begin{itemize}
    \item We introduce a new thermal Trojan that sporadically tampers with SoC's thermal data. Through conducting extensive experiments, we demonstrate that the BIC method fails when confronted with the novel Trojan.
    
    \item We assess SoC vulnerability to thermal Trojans with thermal footprints ranging from 0.1 to 15 degrees Celsius. Existing methods fail to detect thermal attacks with smaller temperature modifications. 
    
    \item We study the capability of tiny machine learning classifiers in detecting thermal Trojans by learning from the steady-state temperature readings.

\end{itemize}
    

\section{Threat model and Impacts}
\label{sec:Background}

This work investigates adversaries capable of compromising DTM systems in mobile SoCs through the insertion or exploitation of Hardware Trojans (HTs). The adversaries can operate across multiple stages of the semiconductor supply chain, including malicious insiders in design or fabrication processes \cite{bhat2017power}, external attackers leveraging untrusted third-party intellectual property (3PIP) cores, and those employing compromised electronic design automation (EDA) tools \cite{sebt2018circuit,10959070,ahmari2025evaluating}. HTs are strategically embedded in thermally sensitive regions, such as near CPU/GPU cores, power management units, or within analog/digital interfaces of thermal sensors \cite{abdelrehim2022bic}. By under-reporting temperatures to disable performance throttling or over-reporting to trigger unnecessary power and clock adjustments, these HTs subvert DTM decision-making logic. Often, such HTs remain dormant until activated by specific thermal profiles (e.g., sustained high-temperature workloads) or external triggers, enabling stealthy, context-aware attacks. Beyond these foundational threats, adversaries may deploy increasingly sophisticated HTs with adaptive capabilities, posing unique challenges to detection and mitigation.

Building on these hardware-level threats, adversaries may deploy adaptive HTs that dynamically alter their activation patterns in response to runtime conditions or defensive measures, enhancing their ability to evade detection. These advanced HTs may leverage machine learning or analog circuitry to mimic benign sensor noise, bypassing static detection mechanisms such as threshold checks or Blind Identification Countermeasures (BIC) \cite{abdelrehim2022bic}. Particularly challenging are analog-based HTs embedded in sensor interfaces, which operate outside digital scan chains and exploit subtle signal distortions to manipulate temperature readings \cite{elahi2024mattermultistageadaptivethermal}.

Such attacks can severely impact SoC performance, reliability, and security. Performance suffers as the DTM system makes incorrect decisions about frequency and core throttling, reducing speed and efficiency, while reliability is compromised by inadequate overheating protection, risking system failures and accelerated aging. Additionally, these manipulations create security vulnerabilities, enabling attackers to exploit the SoC for unauthorized access or operational disruption, threatening user data and device integrity. To counter these risks, implementing solutions like  BIC is vital to detect and mitigate attacks, preserving the SoC's functionality.

\section{iThermTroj: An Intermittent Thermal Trojan}
\label{Sec:Proposed}

According to the literature, the persistence of thermal anomalies injected by an attacker results in a continuous alteration of the temperature readings within the victim core of the target SoC. This sustained manipulation fundamentally challenges the efficacy of existing detection and mitigation mechanisms, including BIC. The consistent nature of thermal attacks allows them to evade detection methods, making the SoC more susceptible to nuanced and stealthy threats. However, thermal attacks can become more sophisticated and stealthy through intermittent thermal alterations. 


\begin{algorithm}[t]
    \caption{ iThermtroj Attack
}
     \begin{algorithmic}[1]
        \STATE $Attack_{Scenario} = Lowering, Elevation, or Fluctuation $
        \STATE $Attack\_Rate = Pick \, from \, \{100\%, 80\%, 60\%, 40\% \}  $
        \STATE $IDX= Randomly \, Chosen \, Core \, Index$
        \FOR{$ steady\_state\_values $} 
            \STATE $random\_value = generate\_random\_number(0, 1)$
            \IF{$random\_value\leq Attack\_Rate$}
                \IF { $T\_error\neq0$ }
                \STATE {$\textbf{Switch} (Attack_{Scenario})$}
                \STATE{$\{$}
                \STATE{$\textbf{CASE} = Lowering : $}
                \STATE {$ T(IDX) = T(IDX) - T_{error}$}
                \STATE {$\textbf{CASE} = Elevation : $}
                \STATE {$ T(IDX) = T(IDX) + T_{error}$}
                \STATE {$\textbf{CASE} = Fluctuation : $}
                \STATE{$T(IDX) = T(IDX) \mp T_{error}$}
                \STATE{$\}$}
                \ENDIF
            \ENDIF
        \ENDFOR            
    \end{algorithmic}
    \label{alg:iThermTroj}
\end{algorithm}

In this section, we introduce a novel and more evasive attack called Intermittent Thermal Trojans (iThermTroj) that works by taking into account SoC's actual thermal traces. This can be done in multiple ways, for example, the attacker can feed the layout information to the HotSpot 6.0 thermal simulator to generate the corresponding thermal traces. The attacker then selects the victim core (can be selected randomly) and injects $\Delta \text{t\_error}$ , which is done by altering the HotSpot's reported temperature. The $\Delta \text{t\_error}$ might be injected to the chosen core sporadically according to \textit{Temperature Lowering}, \textit{Temperature Elevation} or \textit{Temperature Fluctuation} scenarios.

The key point of the proposed iThermTroj attack is that unlike traditional thermal Trojan attack scenarios, it does not involve permanent temperature manipulation. Instead, it follows an intermittent injection process in which some sensor readings are tampered with and the rest of the data points are left untouched. This will help iThermTroj to stay hidden and bypass the most recent detection methods presented in the literature \cite{abdelrehim2022bic}. In this regard, we have considered iThermTroj impacting different percentages of the SoC's thermal traces i.e., scenarios at which iThermTroj is active for 80\%, 60\%, and 40\% of the victim core. The detailed structure of the proposed attack is provided in \Cref{alg:iThermTroj}. In the first two lines of the algorithm, the attacker selects the attacking scenario as well as the attack rate. The attack rate determines the percentage of the thermal data which will be manipulated by iThermTroj. Then, in line 3, the victim core is selected and a random value is generated to be used to manipulate the actual thermal reading according to the selected scenario (lines 10-15).

\begin{figure}[b]
    \centering
    \includegraphics[scale=.25]{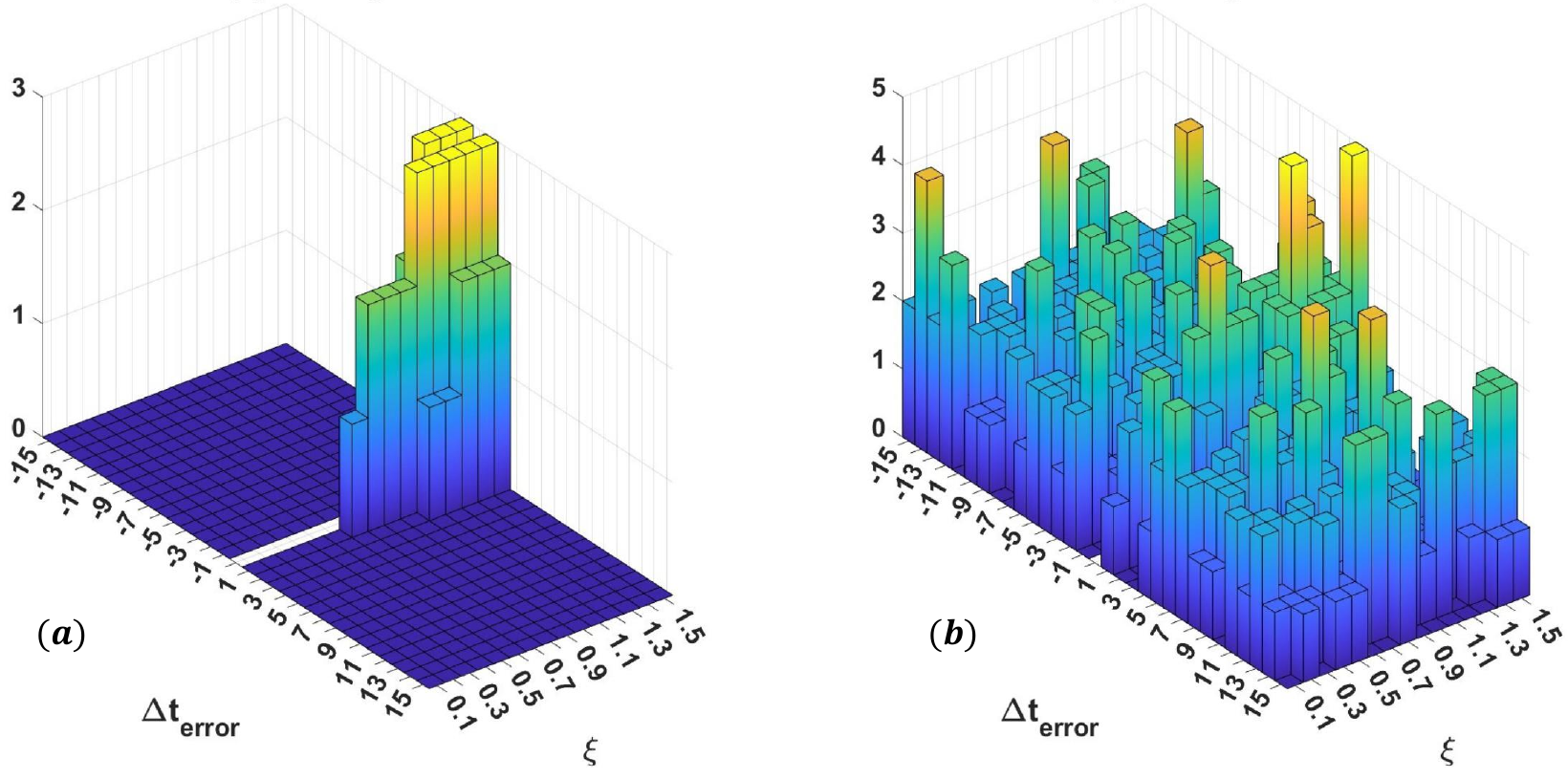}
    \caption{The number of detection failures relative to $\Delta \text{t\_error}$  for a heterogeneous core layout a) with normal attack b) with iThermTroj attack } 
     \label{fig:AttackScenario}
\end{figure}

To highlight this vulnerability, we have performed a detailed analysis comparing the failure rates of BIC under a conventional persistent attack scenario with those observed during our proposed iThermTroj attack. \Cref{fig:AttackScenario}.a and \Cref{fig:AttackScenario}.b compare the efficacy of BIC once encountered persistent and intermittent thermal Trojans respectively. Each plot reports the number of times BIC failed to detect a thermal attack (the lower the better) as a function of various $\Delta \text{t\_error}$ values. The experiment illustrates that under typical persistent attack conditions, BIC demonstrates robust performance, maintaining system integrity for $\Delta \text{t\_error}$ values above $3$ and below $-2$. However, this effectiveness significantly diminishes when the SoC is subjected to the iThermTroj attack, as illustrated in \Cref{fig:AttackScenario}.b. This stark contrast highlights a critical vulnerability, emphasizing the urgent necessity for developing a more resilient countermeasure capable of defending against sophisticated threats like the iThermTroj attack. Enhanced security measures must be prioritized to ensure the SoC's protection and maintain system reliability in the face of evolving attack methodologies.

\section{Proposed Machine Learning Detection}
\label{Sec:countermeasure}


\begin{figure*}[t]
    \centering
    \includegraphics[width=\linewidth]{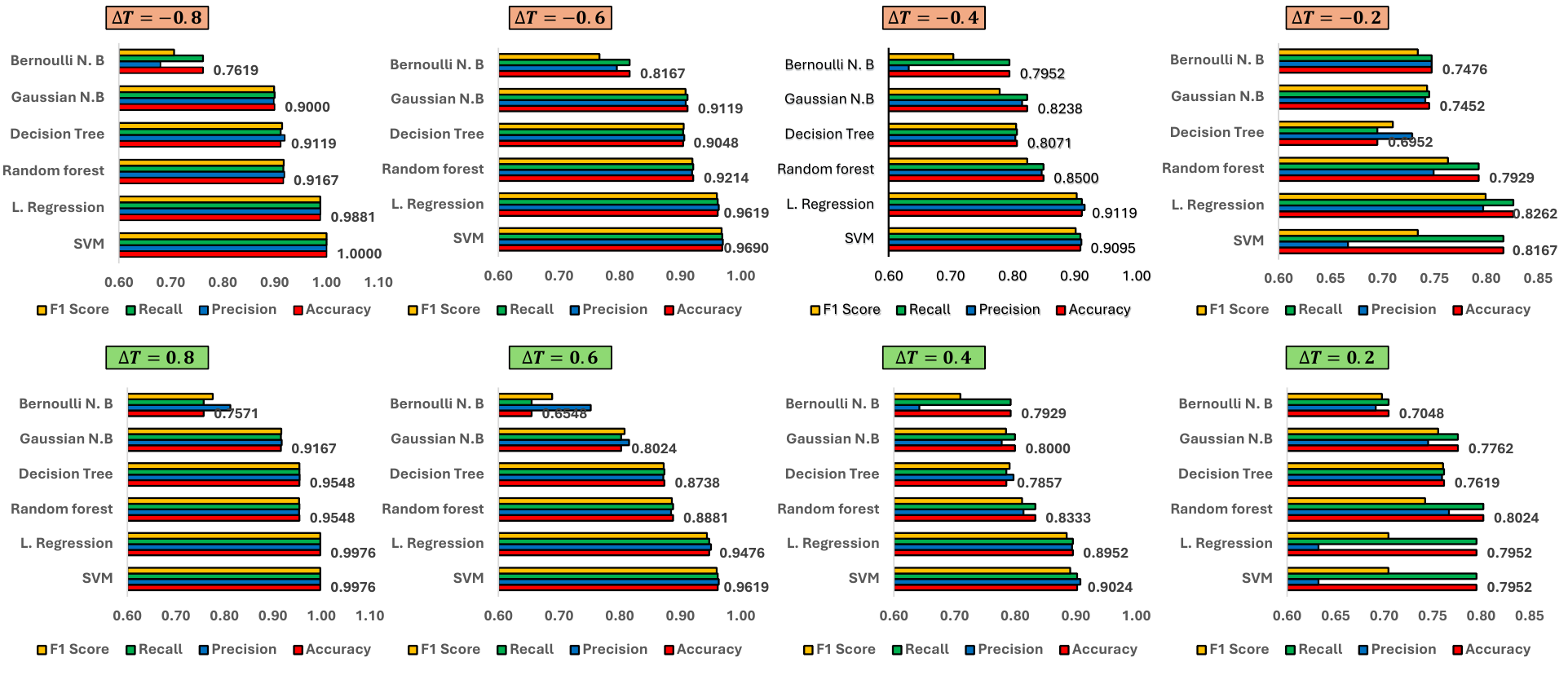}
    \caption{Output parameters of six machine learning classifiers applied to 80\% of the thermal data, tested across $\Delta \text{t\_error}$ intervals from -0.8 to 0.8 with a step size of 0.2. } 
    \label{fig:total}
\end{figure*}

This research addresses limitations in traditional thermal monitoring systems which struggle to detect sophisticated thermal attacks like iThermTroj. TinyML classifiers emerge as a promising solution due to their ability to operate efficiently within mobile devices' resource constraints while processing thermal data in realtime. These classifiers can be deployed directly on SoCs to minimize detection latency and enable continuous monitoring without network dependency. The study evaluates five different classifiers-SVM, Logistic Regression, Random Forest, Decision Tree, and two variants of Naive Bayes-for anomaly detection in thermal data. The methodology involved collecting steady-state temperature readings, deliberately introducing adversarial attacks on 80\% of the data to simulate malicious activity, and using this combined dataset to train and test the models. This approach aims to assess each classifier's effectiveness in identifying thermal anomalies that could indicate attacks, ultimately enhancing thermal management systems while extending device longevity by preventing excessive thermal stress.

\section{Evaluation Results \& Discussions}
\label{Sec:Res}

We utilize a heterogeneous 6-core mobile processor layout as detailed by \cite{gong2017thermal} for our evaluations. For each scenario, we undertake a series of methodical steps. Initially, we employ the HotSpot 6.0 Thermal Simulator \cite{huang2006hotspot} to process the layout information along with the power traces, thereby generating the corresponding thermal traces necessary for our analysis. Subsequently, we act as an attacker by selecting one of the six cores at random to apply a thermal error, denoted as $\Delta \text{t\_error}$ , to the targeted core (intermittently and sporadically). This attack is repeated across different portions of the thermal traces, i.e., 80\%, 60\% and 40\% of the generated thermal traces to assess its impact under varying conditions. We used this data to train used ML classifiers with the distribution of 70\% and 30\% for training and inference. 

\Cref{fig:total} shows the performance of the used ML classifiers in terms of accuracy, recall, precision and F1-score under an iThermTroj attack affecting 80\% of thermal data. Results consistently exhibit the high performance of these classifiers across all evaluated metrics, indicating their robustness and reliability in detecting and counteracting the iThermTroj attack. To extend our analysis we repeated the simulations for 60\% and 40\% thermal data corruption caused by the iThermTroj attack. To account for variations, average precision data across all $\Delta \text{t\_error}$ and injection rates were calculated and visualized in \Cref{fig:AttackCollection}. This figure offers a two-fold interpretation of the classifiers' performance under attack:

\begin{figure*}[t]
    \centering
    \includegraphics[width=\linewidth]{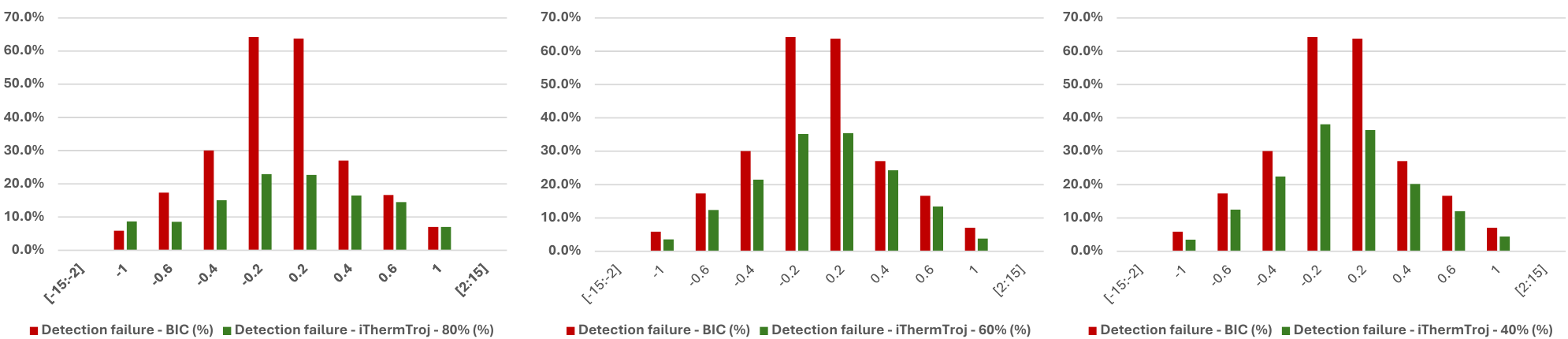}
    \caption{Detection Failure Rates of the ML Countermeasure Across Different Trojan Injection Rates and $\Delta \text{t\_error}$ Thresholds } 
    \label{fig:AttackCollection}
\end{figure*}

\textbf{Accuracy Analysis:} The charts indicate that detection failure rates (i.e., the cases at which ML classifiers failed in detecting iThermTroj attack) increase as the thermal injection becomes smaller, with the BIC method demonstrating greater vulnerability compared to our proposed iThermTroj at various injection rates. Specifically, BIC shows a detection failure rate of 52.38\% under certain conditions. However, when applying an 80\% Trojan injection by iThermTroj attack, the proposed ML detection countermeasure significantly reduces the detection failure rate to approximately 23\%, representing a reduction of around 29\% compared to the highest failure rate observed with the BIC method. Additionally, as attacks become more sophisticated and evasive, the detection failure rate correspondingly rises. For instance, with fault injection rates of 60\% and 40\%, the proposed ML detection countermeasure fails to detect 35.16\% and 38.10\% of attack cases, respectively. Moreover, the distribution form of the chart data appears to follow a normal distribution, which suggests that the variations in detection failure rates are systematically related to the evasiveness of the attacks.
     
\textbf{Resolution Analysis:} The reported results indicate that iThermTroj's ML countermeasure offers a compelling solution, demonstrating not only high accuracy but also superior resolution in terms of $\Delta \text{t\_error}$. Specifically, the BIC method can fully secure the system for $\Delta \text{t\_error}$ values below -2 and above 3. In contrast, even in the worst-case scenario for iThermTroj's ML countermeasure, with a 40\% fault injection rate, we observe complete detection coverage for $\Delta \text{t\_error}$ values below -2 and above 2. This resolution is further enhanced at 60\% and 80\% injection rates, where we achieve complete detection for $\Delta \text{t\_error}$ values below -1.2 and -1, and above 1.2 and 1, respectively. This demonstrates the superior detection capabilities of iThermTroj's ML countermeasure, particularly at higher injection rates, ensuring robust system security across a broader range of $\Delta \text{t\_error}$ values.

\section{Conclusions}
\label{Sec:ConC}

This study presents a significant advancement in the field of thermal management and security for mobile System-on-Chips (SoCs) through the introduction of Intermittent Thermal Trojans (iThermTroj) and the application of machine learning-based anomaly detection techniques. By identifying the vulnerabilities posed by intermittent thermal Trojan attacks and proposing the integration of tiny Machine Learning classifiers, the research offers a novel approach to enhance the resilience and effectiveness of SoC security measures. The findings demonstrate the potential of machine learning models to adapt to complex thermal behaviors and detect anomalies in real-time, thereby improving the responsiveness and longevity of mobile devices. This research contributes valuable insights into mitigating the risks associated with thermal attacks on SoCs and sets a foundation for further advancements in securing modern processors against evolving security threats.

 \newpage
\bibliographystyle{References-Style/IEEEtran}
\bibliography{References-Style/IEEEabrv,References-Style/IEEEexample}
\end{document}